\documentclass{optica-article}

\journal{opticajournal} % for journals or Optica Open

\articletype{Research Article}

\usepackage{lineno}

\usepackage{graphicx}%
\usepackage{multirow}%
\usepackage{amsmath,amsfonts}%
\usepackage{mathrsfs}%
\usepackage[title]{appendix}%
\usepackage{textcomp}%
\usepackage{manyfoot}%
\usepackage{booktabs}%
\usepackage{algorithm}%
\usepackage{algorithmicx}%
\usepackage{algpseudocode}%
\usepackage{listings}%
\usepackage{subcaption}
\usepackage{tabularx}

\usepackage{algpseudocode}

\makeatletter
\renewcommand\OEtitle[1]{%
  {\fontsize{16pt}{18.5pt}\sffamily\bfseries\selectfont\raggedright\textls{#1}\par\vskip.15in}%
}
\let\title\OEtitle
\makeatother

\begin{document}
	
	\title{Intensity-guided pose-free pairwise registration for single-photon view fusion}
	
	\author{Jinyi Liu,\authormark{1} Lijun Liu,\authormark{2*} Shuming Cheng\authormark{1,3**}, Xiaomin Hu\authormark{4,5}, Yiguang Hong\authormark{1,3} and Weiping Zhang,\authormark{1,6***} }
	
	\address{\authormark{1}State Key Laboratory of Autonomous Intelligent Unmanned Systems, Shanghai Research Institute for Intelligent Autonomous Systems, Tongji University, Shanghai, 201203, China\\
        \authormark{2}Department of Mathematics and Computer Science, Shanxi Normal University, Taiyuan 030006, China\\
	 \authormark{4}CAS Key Laboratory of Quantum Information, University of Science and Technology of China, Hefei, 230026, China\\
		\authormark{5}CAS Center For Excellence in Quantum Information and Quantum Physics, University of Science and Technology of China, Hefei, 230026, China\\
        \authormark{3}The Department of Control Science and Engineering, Tongji University, Shanghai, 201804, China\\
        	\authormark{6}The Department of Structural Engineering, Tongji University, Shanghai, 201804, China
	}

	\email{\authormark{*}lljcelia@126.com}
	\email{\authormark{**}shuming\_cheng@tongji.edu.cn} 
	\email{\authormark{***}weiping\_zh@tongji.edu.cn}
	
	\begin{abstract*} 
    Single-photon light detection and ranging (LiDAR) extends active three-dimensional sensing at the fundamental level and has found applications in extreme environments involving long-range operation, low-reflectance targets, and adverse visibility. However, the acquired measurements often give rise to single-photon point clouds that are sparse, spatially non-uniform, and corrupted by outliers and depth distortions, making  pairwise registration challenging especially when sensor poses are not accurately known. In this work, we present a geometry-intensity coupled registration framework (GIC-Reg)  for pose-free pairwise registration in single-photon multi-view sensing. It is established by combining  intensity-guided preprocessing, joint geometry-intensity grid feature aggregation, global matching, and local ambiguity disambiguation to estimate inter-view rigid transformations and hence to support subsequent single-photon view fusion. On the synthetic benchmark, it achieves the lowest relative rotation error (RRE), relative translation error, and root mean square error across all background-noise and dropout rates, in comparison to baselines. Notably, under the most degraded dropout, it reduces the RRE from $13.167^\circ$ to $8.459^\circ$ compared with the learning-based baseline. Furthermore, experimental results on  real two-view data acquired at about 80~m show that it achieves more reliable global orientation and local alignment. Our results show that photon intensity provides an effective  reliability-related cue for stabilizing  pairwise registration in single-photon point clouds, and thus our work  supports practical pose-free registration for single-photon sensing.

	\end{abstract*}
	
	%%%%%%%%%%%%%%%%%%%%%%%%%%  body  %%%%%%%%%%%%%%%%%%%%%%%%%%
	
	\section{Introduction}
	
	Active light detection and ranging (LiDAR) systems have become a cornerstone for acquiring three-dimensional (3D) spatial information in emerging areas such as autonomous driving~\cite{zhang2022large,preussler2019photonically,chang2019new,shi2019photonic,chan2025flash}, remote-sensing mapping~\cite{maeda2025expanding,cheng2017generalized,mao2022polarization,yuan2025remote}, and digital-twin modeling~\cite{scholes2024robust,fan2021unsupervised}. By measuring photon time-of-flight (ToF), they estimate range and reconstruct scene geometry when sufficient return photons are collected under typical illumination and reflectance conditions~\cite{xu2025photon,tachella2019real}. As applications push toward extreme environments, the echo level tends to enter into the photon-starved regime where conventional systems are limited by sensitivity and signal-to-noise ratio~\cite{li2025noise,liu2023compact,lee2023caspi,tobin2021robust}. Therefore, single-photon LiDAR systems have appealed great attention for weak-signal ranging by using single-photon avalanche diodes (SPADs) with picosecond-scale temporal resolution~\cite{Li2021,cheng2024toward,li2024high,Shin2015}. Combined with time-correlated single-photon counting (TCSPC)~\cite{Rayman2006,Zang2019}, they accumulate photon arrival times into time-of-flight histograms and enable depth recovery in extreme environments, such as long range~\cite{Li2020,Li2020a,Li2021,dai2023long,hadfield2023single}, dense fog~\cite{Shi2022,Zhang2022,buller2023single,jiang2023long}, and underwater~\cite{maccarone2015underwater,Maccarone2023,Katzschmann2018,shangguan2023compact}.
	
	Single-photon data are fundamentally governed by stochastic photon-counting statistics which usually suffer from distinctive geometric uncertainty~\cite{Shin2015} and is sensitive to the incidence angle of the probing beam~\cite{Rapp2017,liu2025point,wennoise}. Besides, under oblique illumination or near geometric boundaries, the return-photon time distribution can be broadened substantially and may even be overwhelmed by background counts, leading to non-negligible drift and error in depth estimates. Consequently, the single-view observation is inherently incomplete due to finite field of view, thus motivating multi-view 3D reconstruction to collect complementary geometric features from different observation angles.
	
	One possible approach to multi-view processing of single-photon data is to expand the effective field of view by stitching 2D depth maps and to merge multiple narrow-FOV observations into a panoramic representation~\cite{yang2024pe}. The other assumes known sensor poses and models raw photon-count histograms using neural radiance fields for transient rendering and view synthesis~\cite{malik2023transient,luo2025transientangelo}. These methods require pose estimation as a prerequisite, however, the accurate pose information of single-photon point clouds may be difficult to obtain in extreme environments, leaving pose-free multi-view registration insufficiently addressed.

 To address the above issue, we present an intensity-guided  pose-free pairwise registration framework for subsequent multi-view fusion  in single-photon sensing, by estimating inter-view transformations directly from single-photon point data rather than relying on externally calibrated poses or traditional geometry-only pipelines. Particularly, it follows a geometry-intensity coupled coarse-to-fine registration pipeline tailored to single-photon measurements. First, photon intensity extracted from the photon histogram is assigned to each denoised point as a confidence cue. Then, coarse node descriptors are extracted by a KPConv-based backbone~\cite{thomas2019kpconv} and further aggregated by an XY-grid strategy, so that both local geometry and intensity-related reliability are preserved for coarse correspondence selection. Finally, transformer-based global matching and optimal-transport-based local refinement are used to estimate pairwise rigid transformations, and the aligned views  can be further fused  into a unified 3D reconstruction. Essentially,  it utilizes photon intensity as a reliability-related measurement cue, and together with geometric registration, it is able to distinguish reliable surface returns from pseudo-structures caused by background photons and dark counts.

  Extensive experiments on both synthetic and real-world datasets are conducted to validate the effectiveness of the proposed method. On the synthetic benchmark, it achieves the lowest relative rotation error, relative translation error, and root mean square error across all background-noise and dropout rates, in comparison to Fast-adpgicp~\cite{zhang20234dradarslam} and the learning-based GeoTransformer~\cite{qin2023geotransformer} as baselines. Furthermore, experimental results on  real two-view data  acquired at about 80~m show that it achieves more reliable global orientation and local alignment. These results demonstrate that photon intensity is an effective  reliability-related  cue for  pose-free pairwise registration  and enables more robust 3D fusion in photon-limited scenes.

  The rest of this work is organized as follows. Sec.~\ref{sec2} formulates  pairwise registration as a key step for pose-free multi-view fusion. Sec.~\ref{sec3} presents the geometry-intensity coupled registration framework for this fusion task. Sec.~\ref{sec4} describes the synthetic and real single-photon datasets and evaluation metrics, while Sec.~\ref{sec5} reports the experimental results. Finally, Sec.~\ref{sec6} concludes the paper.
	
\section{Problem formulation: pairwise registration for single-photon multi-view fusion}\label{sec2}\label{sec:problem}

\subsection{Photon counting and histogram}\label{sec2.1}
\label{subsec:histogram}

\begin{figure}[!htb]
	\centering
	\includegraphics[width=9cm]{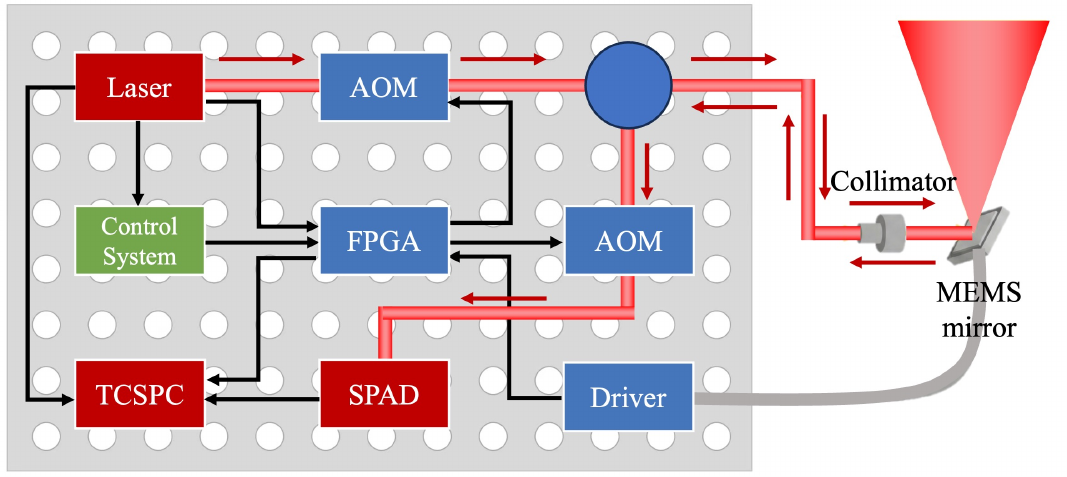}
	\caption{Illustration of the single-photon imaging system.}
	\label{fig1}
\end{figure}

A typical single-photon LiDAR system comprises a pulsed laser source, a SPAD detector, and a TCSPC module~\cite{Shin2015,tachella2019real,hadfield2023single}. During acquisition, the laser emits a sequence of periodic pulses to illuminate the target scene, the SPAD detects returned photons, and the TCSPC module records the ToF information of these collected photons.  Additionally, micro-electro-mechanical systems (MEMS) mirrors switch angles to detect
different pixels, acousto-optic modulators (AOMs) can be integrated to reduce transmission noise, and the FPGA module acts as the central processor to control each component. A complete single-photon imaging system is illustrated in  Fig.~\ref{fig1}. 

In the photon-starved regime where at most one photon is detected within one pulse interval~\cite{Shin2015}, the photon-detection process can be modelled as follows. For each pixel $(u,v)$ with $u=1,\cdots, W$ and $v=1,\cdots, H$, the photon acquisition is carried out over a time horizon $T$ that is evenly divided into $N$ pulse intervals, with interval duration $\Delta = T/N$. Within each interval $n=1,\ldots,N$, a single laser pulse is emitted, the SPAD detects returned photons and records the photon number $a_{uv}[n]$, and the TCSPC records their ToFs described by a vector $\boldsymbol{\tau}_{uv}[n]$.
As at most one photon is detected within one single pulse interval, $\boldsymbol{\tau}_{uv}[n]$ simplifies to a scalar $\tau_{uv}[n]$ so that a pair $(\tau_{uv}[n], a_{uv}[n])$ is indeed recorded during each interval $n$.

The photon-counting histogram can be obtained by choosing a time window and counting falling photons. Specifically, select a time window $[0, T_w]$ with $T_w > \max_n \tau_{uv}[n]$, and discretize it into $K$ bins of width $\delta=T_w/K$. The total photons whose ToFs fall into bin $k$ are counted as
\begin{equation}
h_{uv}[k]=\sum_{n=1,~\tau_{uv}[n]\in [k\delta,(k+1)\delta)}^{N}a_{uv}[n],\quad k=0,\ldots,K-1,  \label{histogram}
\end{equation}
corresponding to the histogram entry. This photon histogram follows approximately a Poisson distribution~\cite{shin2017photon}
\begin{equation}
	h_{uv}(\cdot)\sim \mathcal{P}_{\,N(\eta\,r_{uv}+b_{\mathrm{dc}})}.	\label{eq:counts}
\end{equation}
Here, $\eta\in[0,1]$ is the detection efficiency, $b_{\mathrm{dc}}$ the dark counts, and the photon flux within the $n$-th pulse interval~\cite{shin2017photon}
\begin{equation}
r_{uv}[n]=\int_{n\Delta}^{(n+1)\Delta}\kappa_{uv}\,\rho_{uv}\,s\!\left(t-\frac{2d_{uv}}{c}\right)\,dt+b_{\gamma},\label{forward}
\end{equation}
where $\kappa_{uv}$ describes the radiometric attenuation, $\rho_{uv}$ the surface reflectivity, $s(\cdot)$ the temporal waveform of the emitted pulse, $c$ the speed of light, $b_{\gamma}$ background photons with rate $\gamma$, and $d_{uv}$ the distance range of pixel $(u,v)$. Since the pulse waveform is identical across intervals, $r_{uv}[n]$ is identically distributed as $r_{uv}[n] = r_{uv}$ for all $n$.

\subsection{From photon histogram to single-view point cloud}
\label{sec2.2}

Performing data acquisition over all pixels with $u=1,\dots, W$ and $v=1,\cdots, H$ in a single view yields a histogram tensor
\begin{equation}
\mathbf{H} := \left\{h_{uv}[k] : u=1,\ldots,W;\; v=1,\ldots,H;\; k=0,\ldots,K-1\right\},
\label{data}
\end{equation}
where $h_{uv}[k]$ denotes the photon count at pixel $(u,v)$ in the $k$-th ToF bin. It follows from Eqs.~(\ref{eq:counts}) and~(\ref{forward}) that the imaging task amounts to estimate the depth $d$ of the scene  from this histogram tensor.

Note that the above histogram also contains photon-intensity information~\cite{Shin2015,Rapp2017}. For each pixel $(u,v)$, define its peak bin index as
\begin{equation}
k^{\star}_{uv} = \arg\max_{k\in\{0,\ldots,K-1\}} h_{uv}[k].
\end{equation}
and then a local centroid is computed within a symmetric window of size $w^\prime$ around the peak
\begin{equation}
\bar{k}_{uv} =\frac{\sum_{k\in\mathcal{W}(k^{\star}_{uv};w^\prime)} k\,h_{uv}[k]}{\sum_{k\in\mathcal{W}(k^{\star}_{uv};w^\prime)} h_{uv}[k]},
\label{centroid_local}
\end{equation}
with $\mathcal{W}(k;w^\prime)=\{k-w^\prime,\ldots,k+w^\prime\}\cap\{0,\ldots,K-1\}$. As a consequence, the distance range within each time bin is obtained as 
\begin{equation}
	z_{uv}=\frac{c\,\delta}{2}\,\bar{k}_{uv},
	\label{range_centroid}
\end{equation}
and the intensity attribute, describing the total photon count within the same transient window, is
\begin{equation}
I_{uv} = \sum_{k\in\mathcal{W}(k^{\star}_{uv};w^\prime)} h_{uv}[k].
\label{intensity_local}
\end{equation}

Finally, denote by $\theta_u$ and $\phi_v$ the horizontal and vertical scanning angles, respectively, and the 3D points associated with pixel ($u$,$v$) can be formulated as
\begin{equation}
\mathbf{p}_{uv}=\frac{z_{uv}}{\sqrt{1+\tan^2(\theta_u)+\tan^2(\phi_v)}}\left[\tan(\theta_u),\; \tan(\phi_v),\; 1 \right]^{\top}.
\label{coordinate}
\end{equation}
Together with the photon-intensity information~(\ref{intensity_local}), a raw single-view point cloud is obtained as
\begin{equation}
\mathcal{P}_I := \left\{ (\mathbf{p}_{uv}, I_{uv}) \;\middle|\; I_{uv}>0 \right\},
\label{initialpoint}
\end{equation}
providing a compact representation for the subsequent rigid-body transformation estimation.

\subsection{From single-view to multi-view fusion}
\label{sec2.3}

 Since background photons and dark counts inevitably corrupt the raw dataset $\mathcal{P}_I$ with unstructured outliers~\cite{Shin2015,Rapp2017}, some proper preprocessing technique denoted by $\mathcal{G}(\cdot)$ is usually adopted to suppress noise and to remove spurious points~\cite{tachella2019real}, yielding a denoised yet imperfect point set 
\begin{equation}
\mathcal{P}_D = \mathcal{G}(\mathcal{P}_I).
\label{PD_def}
\end{equation}
However, preprocessing alone cannot eliminate the view dependence of single-photon measurements, because a single view is often incomplete due to the finite field of view and may also contain local distortions induced by photon-counting uncertainty. This motivates  multi-view  fusion that collects complementary geometric features from different observation angles.

When $M$ views of the target scene are collected, it is straightforward to have denoised point sets $\mathcal{P}^{(m)}_D$ with $m=1,\ldots,M$, each of which is expressed in its local sensor frame $\mathcal{F}_m$. Thus, rigid transformations are needed to align all view frames into a global reference frame $\mathcal{F}_0$. For each view $m$, the rigid transformation is described as
\begin{equation}
\mathbf{T}_m= \begin{bmatrix} \mathbf{R}_m & \mathbf{t}_m\\ \mathbf{0}^{\top} & 1 \end{bmatrix} \in \mathrm{SE}(3),
\label{se3}
\end{equation}
where $\mathbf{t}_m\in\mathbb{R}^3$ is a translation vector and $\mathbf{R}_m\in \mathrm{SO}(3)$ is a rotation matrix with $\mathrm{SO}(3)=\{\mathbf{R}\in\mathbb{R}^{3\times 3}\mid \mathbf{R}^{\top}\mathbf{R}=\mathbf{I},\ \det(\mathbf{R})=1\}$. Correspondingly, any local point $\mathbf{p}$ in the view $\mathcal{P}^{(m)}_D$ is mapped to the global frame via
\begin{equation}
\mathbf{p}^{\,\mathrm{g}} = \mathbf{R}_m\mathbf{p}+\mathbf{t}_m.
\label{rigid_map}
\end{equation}
If the pose for each view is known, i.e., the rigid transformation $ \mathbf{T}_m $ is known for $m=1,\cdots, M$, then it is easy to integrate these multi-view data into a unified fusion.

\subsection{Pairwise registration for pose-free multi-view fusion}\label{sec2.4}

 In practice, the accurate pose information is difficult to obtain in single-photon sensing, because it is affected by photon sparsity, dark counts, and view-dependent uncertainty~\cite{Rapp2017,liu2025point}. In this work, we study the pose-free pairwise registration problem, by estimating the inter-view transformation directly from two single-photon point clouds rather than externally calibrating sensor poses.

Particularly, given a source point set $\mathcal{P}^{s}_D$ and a reference point set $\mathcal{P}^{r}_D$ collected from two viewpoints, our aim is to estimate the rigid transformation $\mathbf{T}_{s\rightarrow r}\in \mathrm{SE}(3)$ that aligns the source frame to the reference frame. Define the action of $\mathbf{T}_{s\rightarrow r}$ on the source point set as
\begin{equation}
\mathbf{T}_{s\rightarrow r}(\mathcal{P}^{s}_D) := \left\{ \mathbf{R}_{s\rightarrow r}\mathbf{p}+\mathbf{t}_{s\rightarrow r} \;\middle|\; \mathbf{p}\in\mathcal{P}^{s}_D \right\},
\label{eq:Tm_action}
\end{equation}
and then the pairwise fusion result is written as
\begin{equation}
\mathcal{P}^{\star} = \mathbf{T}_{s\rightarrow r}(\mathcal{P}^{s}_D)\cup \mathcal{P}^{r}_D.
\label{eq:goal}
\end{equation}

Pose-free pairwise registration is challenging in single-photon sensing~\cite{tachella2019real,liu2025point}. First, the point sets are sparse and spatially non-uniform. Second, background photons and dark counts introduce outliers and pseudo-structures. Third, the overlaps between two views can be partial and local geometric ambiguity may lead to incorrect correspondences. Therefore, a reliable pairwise registration strategy should suppress noise, exploit stable structural information, and estimate inter-view transformations robustly.

\begin{figure}[t]
	\centering
	\includegraphics[width=1\linewidth]{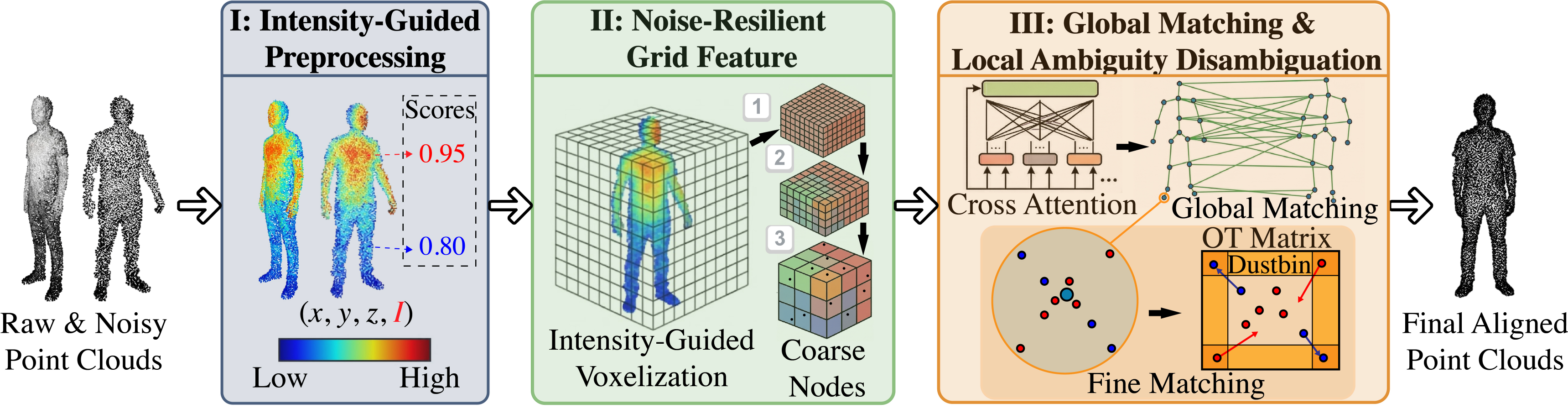}
	\caption{The working pipeline of the geometry-intensity coupled registration framework (GIC-Reg) for raw and noisy point clouds.  The first stage converts photon intensity into confidence scores to indicate point reliability.  The second performs joint geometry-intensity grid feature aggregation to extract coarse geometric nodes and to suppress dark-count noise. The final combines self- and cross-attention with optimal transport (OT) matching to establish global correspondences and refines the final alignment using both geometric and photometric cues.}
	\label{fig:overall}
\end{figure}

\section{Methodology:  geometry-intensity coupled registration framework}\label{sec3}

Here, we propose a geometry-intensity coupled registration framework (GIC-Reg)  for pose-free pairwise registration  of single-photon point clouds. As shown in Fig.~\ref{fig:overall}, it consists of three stages: the first stage is  Intensity-Guided  Confidence Preprocessing, which assigns photon intensity as a confidence cue to each point; the second is Joint Geometry-Intensity Grid Feature Aggregation, which extracts coarse node descriptors with a KPConv-based backbone and further aggregates them into XY-grid cell descriptors; and the final stage is Global Matching and Local Ambiguity Disambiguation, which performs transformer-based coarse correspondence reasoning and OT-based patch refinement for rigid transformation estimation.

\subsection{Intensity-Guided Preprocessing}\label{sec:feature_encoding}

 Note first that in single-photon point clouds, photon intensity is closely related to point reliability, in the sense that weaker returns are more susceptible to background photons and dark counts, whereas stronger returns are usually more stable~\cite{Rapp2017,Shin2015}. Meanwhile, sparse background events may also generate points whose spatial locations are geometrically similar to those of valid surface returns. If only spatial coordinates are used, then noisy points could be interpreted as plausible local structures. Thus, the first stage of GIC-Reg is to incorporate photon intensity as a reliability-related confidence cue for each point in the single-photon point cloud.

Specifically, any point in each denoised point set $\mathcal{P}^{(m)}_D$ in Eq.~(\ref{PD_def}) is assigned with a scalar confidence feature
\begin{equation}
c_i = \mathcal{N}(I_i),
\label{eq:initial_feature}
\end{equation}
 where $I_i$ is given by Eq.~(\ref{intensity_local}) and $\mathcal{N}(\cdot)$ denotes a maximum-based normalization within each scan that rescales photon intensity into a relative confidence value between 0 and 1. The resulting feature $c_i$ is attached to each point and propagated to the subsequent geometry-intensity feature aggregation and matching modules.

\subsection{Joint Geometry-Intensity Grid Feature Aggregation}\label{sec:coarse_matching}

\begin{figure}[htbp]
	\centering
	\includegraphics[scale=0.3]{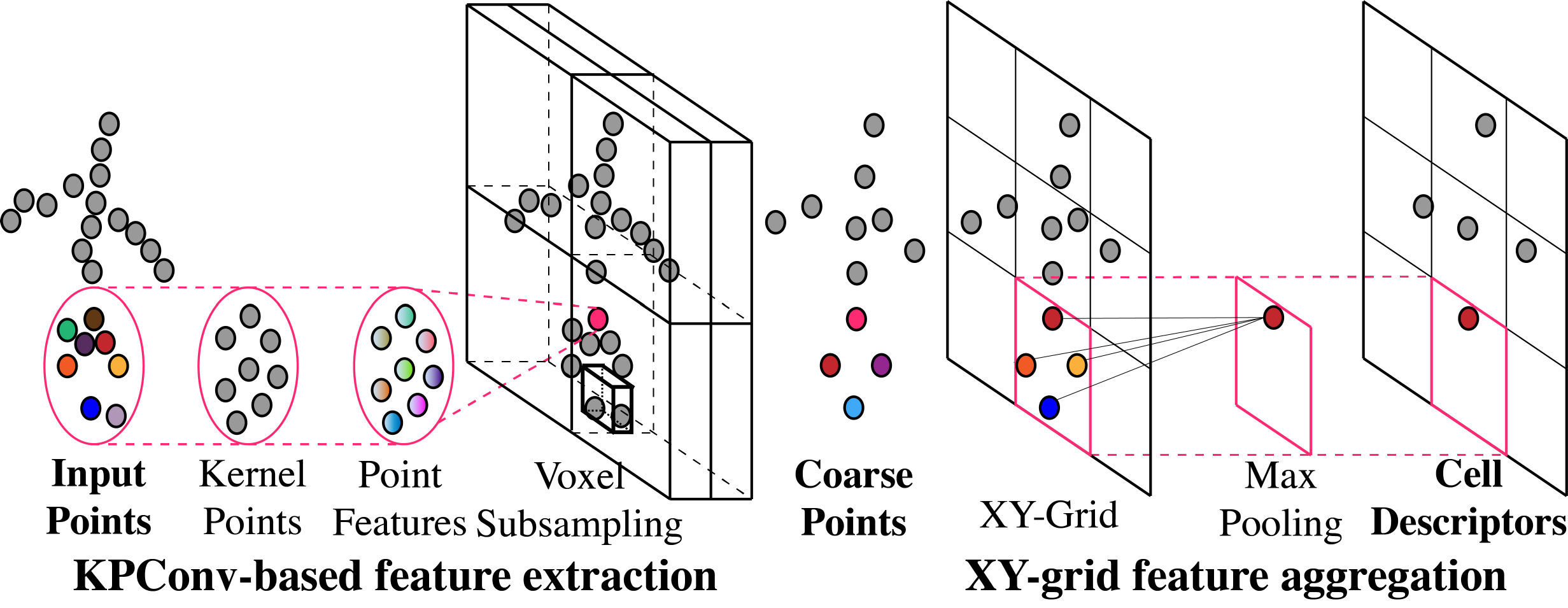}
	\caption{The Joint Geometry-Intensity Grid Feature Aggregation module: All input points are first encoded into coarse nodes and descriptors with a KPConv-based backbone, which are then grouped by planar XY-grid cells and aggregated by max pooling to form cell descriptors for coarse correspondence selection.}
	\label{fig:xygrid}
\end{figure}

As shown in Fig.~\ref{fig:xygrid}, the second stage is configured to extract coarse node descriptors with a KPConv-based backbone~\cite{thomas2019kpconv} and further aggregates these nodes with an XY-grid aggregation strategy.
When $M$ views are involved and denoised point sets $\{\mathcal{P}^{(m)}_D\}_{m=1}^{M}$ are collected, these point clouds are first processed by the KPConv-based backbone, which performs progressive neighborhood aggregation and subsampling to generate coarse nodes and their descriptors. Correspondingly, the coarse node sets are generated as
\begin{equation}
\mathcal{P}^{(m)}_c = \{\mathbf{p}^{(m)}_i\}_{i=1}^{N_m},\label{coarse}
\end{equation}
where the node coordinate $\mathbf{p}^{(m)}_i$ is a geometric representative produced by the subsampling process and $N_m$ is the number of coarse nodes in view $m$, and the  coarse descriptor sets are
\begin{equation}
\mathbf{F}^{(m)}_c = \{\mathbf{f}^{(m)}_i\}_{i=1}^{N_m},
\end{equation}
 where the descriptors $\mathbf{f}^{(m)}_i$ are learned by the KPConv-based backbone using the scalar confidence $c_i^{(m)}$ in Eq.~(\ref{eq:initial_feature}) as the point feature, while the 3D coordinates define the neighborhood support. This process jointly preserves geometric layout and intensity-aware feature information.

In the scanning system as outlined in Fig.~\ref{fig1}, the XY coordinates are determined by MEMS-controlled angular sampling, whereas the depth coordinate is more strongly affected by photon-counting uncertainty. Therefore, the planar constraint that the coarse nodes are grouped into planar grid cells based on their XY coordinates is further imposed before coarse correspondence selection. Given any coarse node set $\mathcal{P}^{(m)}_c$ as per Eq.~(\ref{coarse}), a grid-cell set is obtained as
\begin{equation}
\mathcal{G}^{(m)}=\{g^{(m)}_u\}_{u=1}^{U_m},
\end{equation}
 where the grid is constructed from the reconstructed Cartesian $x$-$y$ coordinates in each local sensor frame, each $g^{(m)}_u$ denotes the subset of coarse nodes falling into the $u$-th grid cell, and empty cells are excluded so that $U_m$ is the number of non-empty cells. The default grid-cell size is 0.15, with its sensitivity analyzed in Supplement 1.  In each non-empty cell, the coarse descriptors are aggregated by max pooling 
\begin{equation}
\mathbf{z}^{(m)}_u = \max_{\,i:\mathbf{p}^{(m)}_i \in g^{(m)}_u} \mathbf{f}^{(m)}_i,
\end{equation}
where the max operator is element-wise over the descriptors.  This grid max-pooling strategy aggregates the learned coarse descriptors within each XY cell so that the dominant and more reliable responses become more prominent. Compared with average pooling, max pooling is less likely to dilute sparse high-confidence responses by mixing them with noisy or weak-return descriptors in the same cell. Compared with attention-based pooling, it introduces no additional trainable parameters and is therefore less prone to overfitting under the limited single-photon training data.  As a consequence, the cell descriptor $\mathbf{z}^{(m)}_u$ provides a joint geometry-intensity representation and preserves both the intensity prior and the geometric scale information.

\subsection{Global Matching and Local Ambiguity Disambiguation}\label{sec:ot_matching}

The final stage first employs a geometric transformer to jointly process the coarse nodes $\mathcal{P}^{(m)}_c$, coarse descriptors $\mathbf{F}^{(m)}_c$, and XY-grid cell descriptors $\mathbf{z}^{(m)}_u$ as inputs, for global correspondence reasoning, and then formulates the rigid transformation estimation as regularized OT based on the XY-grid representation.

The transformer operates that the geometry branch uses the spatial coordinates of coarse nodes to compute the relative distances and angles between nodes and hence to encode them into a relative positional embedding $\mathbf{p}_{ij}$, while the feature branch operates on coarse descriptors under the confidence cue guidance to encode the intensity-aware information together with local geometric structure. Specifically, if the descriptor at node $i$ is projected to the query vector $\mathbf{q}_i$ and  the descriptor at node $j$ is projected to the key $\mathbf{k}_j$, then the attention score between nodes $i$ and $j$ is computed as
\begin{equation}
A_{ij} = \frac{\mathbf{q}_i^{\top}\mathbf{k}_j + \mathbf{q}_i^{\top}\mathbf{p}_{ij}}{\sqrt{d}}
\label{eq:attention_score}
\end{equation}
with the feature dimension $d$. Here, $\mathbf{q}_i^{\top}\mathbf{k}_j$ measures the similarity of learned coarse descriptors, and $\mathbf{q}_i^{\top}\mathbf{p}_{ij}$ quantifies the geometric prior through the relative positional embedding. Thus, the updated coarse descriptors jointly encode geometry-aware spatial arrangement and intensity-influenced feature similarity.

After transformer interaction, the fine correspondence of each matched coarse node pair is estimated by regularized OT inspired by GeoTransformer~\cite{qin2023geotransformer}, with a dustbin introduced to reject noisy or unmatched  single-photon points. The XY-grid branch is used to regularize coarse candidate selection and supervision, while the final rigid transformation $\mathbf{T}_{m \to 1}$ is still estimated from the refined correspondences by weighted singular value decomposition. This cell-guided strategy reduces incorrect coarse connections caused by noisy local structures and repeated low-resolution patterns, while preserving the informative descriptors learned by the backbone and transformer.

\subsection{Multiscale joint optimization objective}\label{sec:loss_function}

A multiscale supervision scheme is constructed jointly on the coarse descriptor level and the fine correspondence level to train the above modules. At the coarse level, the loss $\mathcal{L}_{\mathrm{coarse}}$ is imposed on the transformer-updated coarse descriptors after transformer interaction, and at the fine level, the loss $\mathcal{L}_{\mathrm{fine}}$ is imposed on the OT transport matrix for local patches. It supervises the transport assignments so that points belonging to true inlier correspondences receive high matching probability, while outliers or unmatched points are assigned to the dustbin.

Thus, the overall objective is defined as
\begin{equation}
\mathcal{L} = \lambda_c \mathcal{L}_{\mathrm{coarse}} + \lambda_f \mathcal{L}_{\mathrm{fine}},
\label{eq:total_loss}
\end{equation}
 where $\lambda_c$ and $\lambda_f$ are balancing weights and are both set to 1.0 in all experiments. The coarse term improves the separability of the learned coarse descriptors, and the fine term improves the reliability of the OT-based local matching.  This joint optimization is designed to enable the network to learn geometry-intensity coupled descriptors and to produce more stable correspondences for rigid transformation estimation.

	\section{Multi-view single-photon datasets}\label{sec4}

\subsection{Synthetic single-photon dataset}\label{sec:synthetic_dataset}

\begin{figure}[htbp]
    \centering
    \includegraphics[scale=0.4]{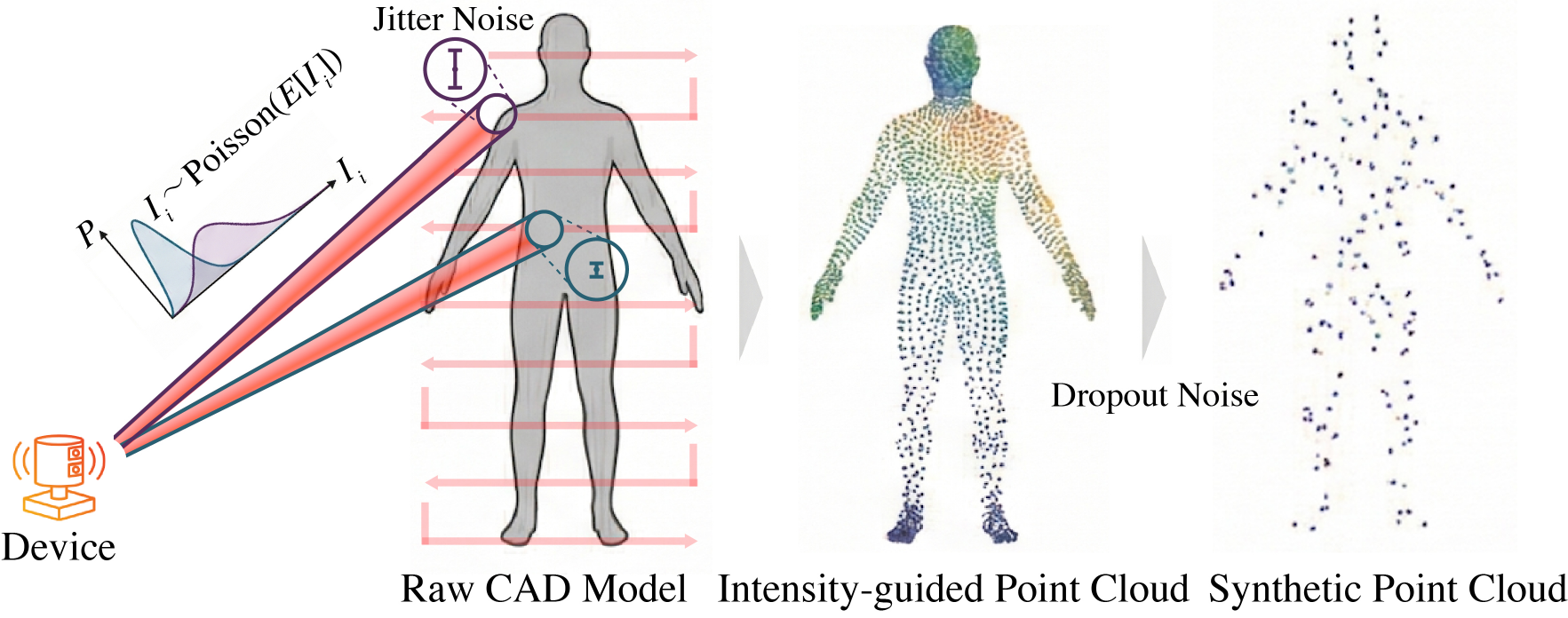}
    \caption{{Synthetic single-photon multi-view dataset generation. The synthetic process models the intensity-dependent degradation, including depth jitter noise and dropout.}}
    \label{synthetic_dataset}
\end{figure}

 A multi-view single-photon dataset is synthesized from ModelNet40~\cite{wu2015modelnet}, as illustrated in Fig.~\ref{synthetic_dataset}. To approximately simulate the photon-starved acquisition process described in Sec.~\ref{sec2.1}, the synthetic data are generated under a measurement-motivated assumption that the expected photon intensity is primarily determined by the effective surface area covered by the distant laser footprint and is further perturbed by photon-counting fluctuations. The footprint scale is set according to the measured beam size of the collimated system. Surface reflectivity and incidence-angle dependence are not explicitly modeled, because the benchmark is intended to provide a data-driven registration test under generic single-photon degradation rather than a material-specific radiometric simulation.

Specifically, photon counts are sampled from a Poisson distribution according to the photon-counting model in Sec.~\ref{sec2.1}. The ideal range along each probing ray is then perturbed by a noise distribution whose variance is inversely related to the intensity $I_i$. To match the operating characteristics of our device, five background-noise levels are considered by scaling the nominal noise setting with factors of $0.5$, $1.0$, $1.5$, $2.0$, and $2.5$, thus forming five groups from mild to severe degradation. In addition, stochastic ray-level dropout is applied to emulate missing photon returns, with the dropout rate varying from $0$ to $0.5$. Background-like outliers are further introduced before the same preprocessing step is applied. This degradation setting reproduces the intensity-dependent depth uncertainty and missing-return behavior observed in the real single-photon LiDAR.

Virtual cameras are placed at different viewpoints to generate source-reference pairs $(\mathcal{P}^s, \mathcal{P}^r)$ together with the corresponding ground-truth relative poses $\mathbf{T}^{\ast}_{s\rightarrow r} \in \mathrm{SE}(3)$ for training and evaluation. The synthetic point clouds are generated at a metric scale consistent with the real acquisition setting, and the scanning resolution, angular sampling pattern, and range setting follow the MEMS-based acquisition geometry. The relative viewpoint change is randomly sampled in the range of $10^\circ$--$30^\circ$, with the target center kept within the field of view to maintain partial overlap between the source and reference scans. The resulting benchmark contains 12000 training pairs, 1500 validation pairs, and 1500 testing pairs. To avoid training-test leakage, the split is performed by CAD category rather than by generated pair, so no CAD object instance is shared between training and testing.

\subsection{Real single-photon dataset}\label{sec:real_dataset}

\begin{figure}[htbp]
    \centering
    \includegraphics[scale=0.55]{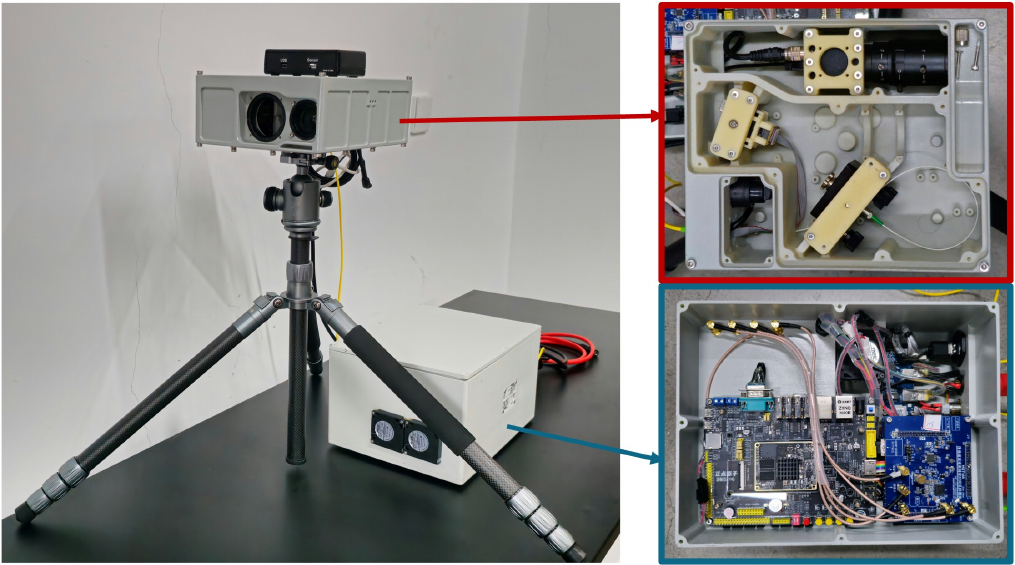}
    \caption{{Schematic diagram and photograph of the coaxial single-photon LiDAR system.}}
    \label{fig:setup}
\end{figure}

A real-world single-photon dataset is also collected from using the coaxial single-photon LiDAR system shown in Fig.~\ref{fig:setup}. The transmission module employs a 1550~nm pulsed fiber laser with a repetition rate of 1~MHz and a pulse width of approximately 1.5~ns. Two AOMs are inserted into the optical path to suppress amplified spontaneous emission noise and to improve the signal-to-noise ratio. The beam is then expanded and collimated before raster scanning by a dual-axis MEMS mirror. On the receiver side, a commercial free-running InGaAs/InP single-photon detector (Quantum CTek QCD600C) is used for detection. The detector provides a detection efficiency of at least 13\% at 1550~nm and a dark count rate below 4~kHz. Photon arrival times are digitized by a TCSPC module. The measured instrument response function  has a full width at half maximum of approximately 1.5~ns, corresponding to an effective root-mean-square timing jitter of about 600~ps under a quasi-Gaussian approximation.  The real data are used only for testing the trained model; they are not included in training or fine-tuning, and no domain-adaptation technique~\cite{chen2022deep} is applied. 

 In the real experiment, the human target had a characteristic size of approximately 80~cm and remained stationary during acquisition. The measurements were acquired sequentially under an underground parking-garage condition. Each scan used a $64\times64$ angular sampling grid, and the acquisition time was 0.2~s per frame. The average detected photon count was approximately 15 photons per pixel. The two views were acquired from different viewpoints with an approximate angular change of about $20^\circ$, which was not used as an input pose. The point cloud was reconstructed by assigning the recovered depth and photon-intensity value of each pixel to the corresponding MEMS scanning ray, followed by the same denoising preprocessing before registration. 

\begin{figure}[t]
    \centering
    \includegraphics[width=1\linewidth]{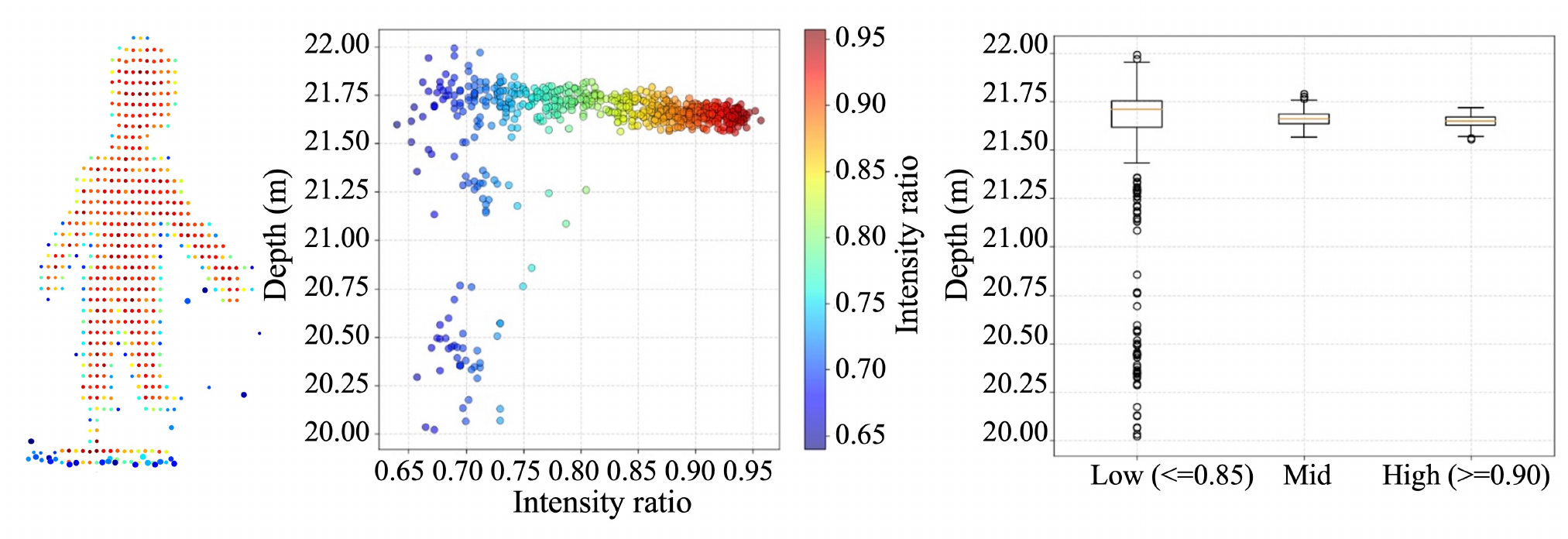}
    \caption{Intensity-dependent depth uncertainty in the real single-photon dataset. Left: the depth map of target. Middle: the scatter plot of intensity versus recovered depth for a representative target range. Right: the depth distributions across different intensity intervals. It demonstrates that the lower-intensity returns exhibit a larger depth variance, while the higher-intensity returns are more narrowly concentrated.}
    \label{fig:intensity_depth_relation}
\end{figure}

 To verify that photon intensity is related to depth stability in real acquisition, Fig.~\ref{fig:intensity_depth_relation} shows the relationship between photon intensity and recovered depth for a representative target range. Lower-intensity returns exhibit a larger depth spread, whereas higher-intensity returns are more concentrated, supporting the intensity-dependent uncertainty assumption used in the synthetic benchmark.

\subsection{Evaluation Metrics}\label{sec:metrics}

 Since reliable ground-truth relative poses are not available for the real long-range single-photon measurements~\cite{malik2023transient}, quantitative pose errors are reported on the synthetic benchmark, while the real experiment is used as a qualitative failure-mode comparison under actual acquisition conditions. In the synthetic benchmark, relative rotation error and relative translation error are used as the primary pose-accuracy metrics because the ground-truth relative transformations are available, while the root mean square error is reported as an auxiliary post-alignment residual for describing geometric consistency after registration.  In particular, denote by $\mathbf{T}^{\ast}_{s\rightarrow r}=(\mathbf{R}^{\ast},\mathbf{t}^{\ast})$ the ground-truth transformation and $\hat{\mathbf{T}}_{s\rightarrow r}=(\hat{\mathbf{R}},\hat{\mathbf{t}})$ the estimated transformation~\cite{qin2023geotransformer}.

\paragraph{Relative rotation error (RRE).}
It is defined as the geodesic distance on ${\rm SO}(3)$ as
\begin{equation}
\mathrm{RRE}=\arccos\!\left(\frac{\mathrm{trace}\!\left((\hat{\mathbf{R}})^{\top}\mathbf{R}^{\ast}\right)-1}{2}\right)\cdot \frac{180}{\pi}.
\end{equation}

\paragraph{Relative translation error (RTE).}
It quantifies the Euclidean distance between the estimated and ground-truth translations as
\begin{equation}
\mathrm{RTE}=\left\|\hat{\mathbf{t}}-\mathbf{t}^{\ast}\right\|_2,
\end{equation}
with the 2-norm $\|\bullet\|_2$ for vectors.

\paragraph{Root mean square error (RMSE).}
 As an auxiliary residual metric, it measures the geometric consistency after registration by computing the root-mean-square nearest-neighbor residual from the transformed source cloud to the reference as 

\begin{equation}
\mathrm{RMSE}=
\sqrt{
\frac{1}{|\mathcal{P}^s|}
\sum_{\mathbf{p}\in\mathcal{P}^s}
\left\|
\hat{\mathbf{R}}\mathbf{p}+\hat{\mathbf{t}}
-\mathrm{NN}\!\left(\hat{\mathbf{R}}\mathbf{p}+\hat{\mathbf{t}}, \mathcal{P}^r\right)
\right\|_2^2
},
\end{equation}
where $\mathbf{p}$ denotes a source point in $\mathcal{P}^s$, and $\mathrm{NN}(\cdot,\mathcal{P}^r)$ denotes the nearest neighbor in $\mathcal{P}^r$.

\section{Experimental results}\label{sec5}

In this section, the proposed GIC-Reg model for  pose-free pairwise registration  is extensively tested on the synthetic and real single-photon datasets. On the synthetic, its registration performance is evaluated under two representative degradations, namely background noise and spatial dropout. On the real, it is examined whether estimated registrations preserve the correct global orientation and the local body-part alignment under sparse photon returns. Fast-adpgicp~\cite{zhang20234dradarslam} is used as a local registration baseline, while GeoTransformer~\cite{qin2023geotransformer} serves as a learning-based one.  For a fair comparison, all registration methods are evaluated on the same preprocessed point sets. Additional comparisons with classical and robust geometry-only registration baselines are provided in Supplement 1, where the numerical mean/median errors and the corresponding failure modes are summarized. 

\subsection{Results on Synthetic Data}\label{sec:synthetic_results}

\subsubsection{Pairwise registration on the synthetic dataset}

\begin{figure}[t]
    \centering
    \includegraphics[width=1\linewidth]{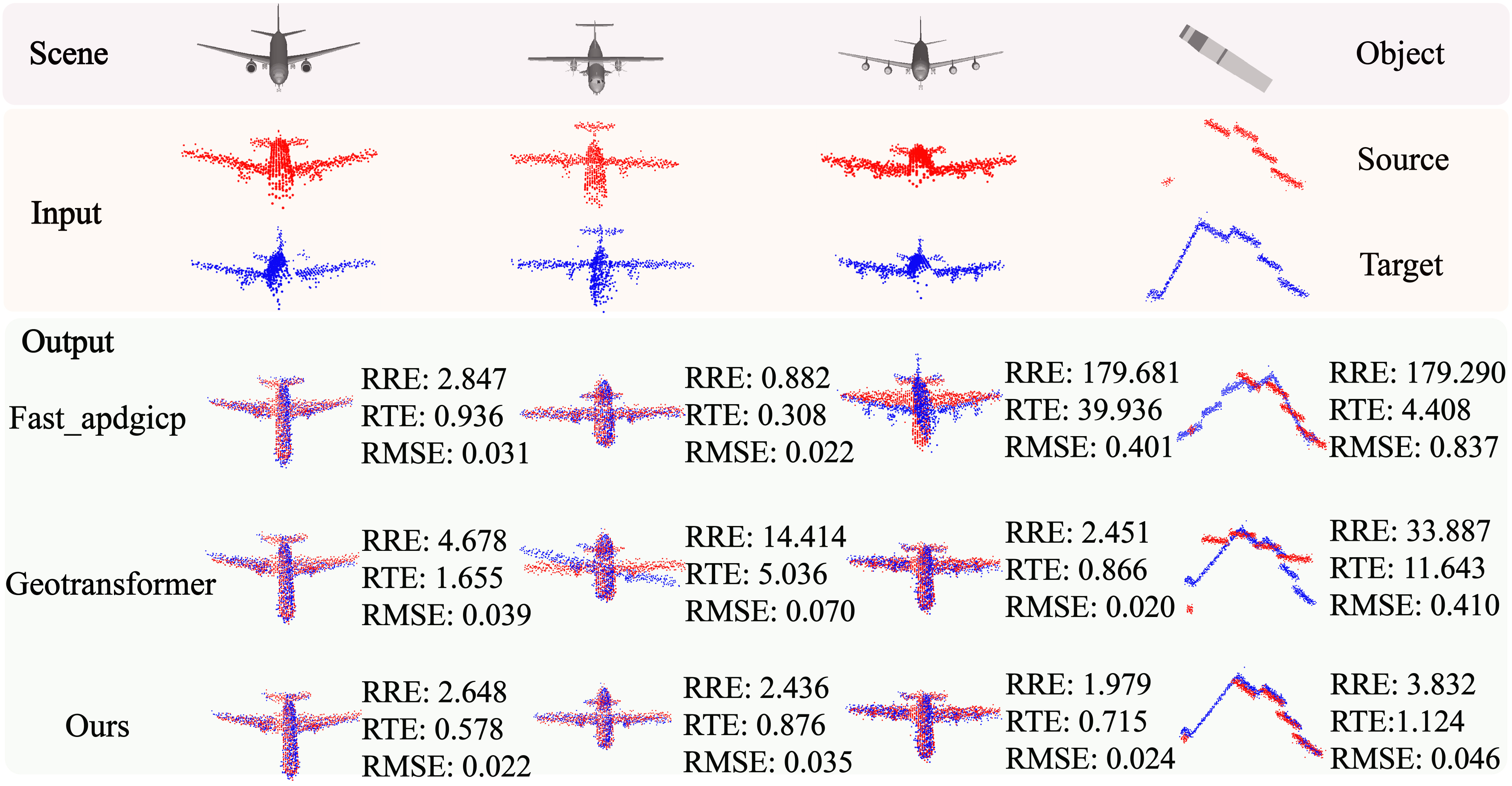}
    \caption{The registration performance on the synthetic datasets. It shows that Fast-adpgicp~\cite{zhang20234dradarslam} can converge to a wrong local optimum in noisy aircraft-body regions, while GeoTransformer~\cite{qin2023geotransformer} aligns part of the structure. Our GIC-Reg model significantly suppresses these false overlaps and successfully recovers the correct global alignment.}
    \label{fig:synthetic}
\end{figure}

The registration performance on the synthetic datasets are shown in Fig.~\ref{fig:synthetic}. Our model successfully recovers the correct global alignment, even with depth perturbations and missing returns, noting that the registered source and reference point clouds overlap more consistently at the source and target. By contrast, Fast-adpgicp converges to an incorrect alignment so that the source cloud is attracted to noisy regions (row 3 and 4), while GeoTransformer heavily relies mainly on geometric coordinates when intensity is not provided, so some part structure can appear aligned while the global pose remains biased toward noise-dominated regions (row 2 and 4). Since local geometric fitting and geometry-only global matching are both vulnerable to single-photon degradation, our method introduces the photon intensity feature into the correspondence learning to reduce this ambiguity and hence to improve the registration performance.

\subsubsection{Robustness against the background noise}

 The registration results under the varying background noise conditions are summarized in Table~\ref{tab:sbr_noise}. It is found that the proposed method maintains lower rotation error, translation error, and alignment residual than both Fast-adpgicp and GeoTransformer. In Group~1, the RRE is reduced from $17.775^\circ$ for Fast-adpgicp and $6.564^\circ$ for GeoTransformer to $4.570^\circ$, while the RTE is reduced to $1.573$~cm and the RMSE is reduced to $0.027$~m. In the most degraded setting, Group~5, it still achieves a lower RRE of $8.517^\circ$ and RMSE of $0.039$~m, while $49.805^\circ$ and $0.203$~m for Fast-adpgicp and $9.109^\circ$ and $0.042$~m for GeoTransformer. The median errors further show that the proposed method remains stable for typical test pairs, while the larger gap between mean and median errors of Fast-adpgicp indicates that its average performance is more affected by occasional severe failures.  It indicates that the photon intensity cue helps suppress rotationally ambiguous matches and keeps the global orientation more stable across different background-noise groups, thus improving the reliability of coarse correspondence selection under noisy single-photon returns. 

\begin{table}[htbp]
  \centering
  \caption{Registration performance under the varying background noise on the synthetic single-photon benchmark. Mean and median errors are reported. }
  \label{tab:sbr_noise}
  \small
  {
  \setlength{\tabcolsep}{3pt}
  \begin{tabular}{llcccccc}
    \toprule
    \textbf{Condition} & \textbf{Method} & \multicolumn{2}{c}{\textbf{RRE ($^\circ$) $\downarrow$}} & \multicolumn{2}{c}{\textbf{RTE (cm) $\downarrow$}} & \multicolumn{2}{c}{\textbf{RMSE (m) $\downarrow$}} \\
    \cmidrule(lr){3-4}\cmidrule(lr){5-6}\cmidrule(lr){7-8}
    & & \textbf{mean} & \textbf{median} & \textbf{mean} & \textbf{median} & \textbf{mean} & \textbf{median} \\
    \midrule
    \multirow{3}{*}{Group 1}
    & Fast-adpgicp & 17.775 & 3.787 & 4.010 & 1.261 & 0.047 & 0.024 \\
    & GeoTransformer & 6.564 & 1.999 & 2.256 & 0.709 & 0.034 & 0.018 \\
    & Ours & \textbf{4.570} & \textbf{1.832} & \textbf{1.573} & \textbf{0.442} & \textbf{0.027} & \textbf{0.010} \\
    \midrule
    \multirow{3}{*}{Group 2}
    & Fast-adpgicp & 17.641 & 3.231 & 3.945 & 1.057 & 0.050 & 0.024 \\
    & GeoTransformer & 8.515 & 2.433 & 2.940 & 0.835 & 0.040 & 0.020 \\
    & Ours & \textbf{7.671} & \textbf{1.479} & \textbf{2.626} & \textbf{0.390} & \textbf{0.037} & \textbf{0.014} \\
    \midrule
    \multirow{3}{*}{Group 3}
    & Fast-adpgicp & 33.086 & 1.949 & 7.076 & 0.610 & 0.084 & 0.022 \\
    & GeoTransformer & 9.476 & 1.466 & 3.238 & 0.508 & 0.041 & \textbf{0.018} \\
    & Ours & \textbf{8.397} & \textbf{1.435} & \textbf{2.888} & \textbf{0.478} & \textbf{0.038} & \textbf{0.018} \\
    \midrule
    \multirow{3}{*}{Group 4}
    & Fast-adpgicp & 49.662 & 4.737 & 9.573 & 1.665 & 0.152 & 0.026 \\
    & GeoTransformer & 11.272 & 2.987 & 3.890 & 0.999 & 0.049 & 0.022 \\
    & Ours & \textbf{8.696} & \textbf{1.329} & \textbf{2.988} & \textbf{0.386} & \textbf{0.039} & \textbf{0.020} \\
    \midrule
    \multirow{3}{*}{Group 5}
    & Fast-adpgicp & 49.805 & 4.511 & 7.756 & 1.477 & 0.203 & 0.027 \\
    & GeoTransformer & 9.109 & 2.387 & 3.127 & 0.830 & 0.042 & 0.021 \\
    & Ours & \textbf{8.517} & \textbf{1.379} & \textbf{2.934} & \textbf{0.364} & \textbf{0.039} & \textbf{0.014} \\
    \bottomrule
  \end{tabular}
  }
\end{table}

\subsubsection{Robustness against the spatial sparsity}

 The comparative results on the varying spatial dropout are summarized in Table~\ref{tab:dropout}. In comparison to Fast-adpgicp and GeoTransformer, our method consistently admits smaller RRE, RTE, and RMSE. At the $10\%$ dropout, the RRE decreases from $24.929^\circ$ for Fast-adpgicp and $8.320^\circ$ for GeoTransformer to $4.743^\circ$, while the RTE and RMSE are reduced to $1.637$~cm and $0.028$~m, respectively. At the $50\%$ dropout, the proposed method still achieves a lower RRE of $8.459^\circ$ and RMSE of $0.052$~m, compared with $74.038^\circ$ and $0.196$~m for Fast-adpgicp, and $13.167^\circ$ and $0.057$~m for GeoTransformer. The mean--median difference is especially visible for Fast-adpgicp at high dropout rates, suggesting that sparse geometric support can lead to occasional severe failures. In contrast, the proposed method keeps both mean and median pose errors lower, indicating more stable behavior across test pairs.  These results confirm that our method remains effective when the geometric support becomes sparse.

\begin{table}[htbp]
  \centering
  \caption{Registration performance under the varying dropout rates on the synthetic single-photon benchmark. Mean and median errors are reported. }
  \label{tab:dropout}
  \small
  {
  \setlength{\tabcolsep}{3pt}
  \begin{tabular}{llcccccc}
    \toprule
    \textbf{Dropout Rate} & \textbf{Method} & \multicolumn{2}{c}{\textbf{RRE ($^\circ$) $\downarrow$}} & \multicolumn{2}{c}{\textbf{RTE (cm) $\downarrow$}} & \multicolumn{2}{c}{\textbf{RMSE (m) $\downarrow$}} \\
    \cmidrule(lr){3-4}\cmidrule(lr){5-6}\cmidrule(lr){7-8}
    & & \textbf{mean} & \textbf{median} & \textbf{mean} & \textbf{median} & \textbf{mean} & \textbf{median} \\
    \midrule
    \multirow{3}{*}{$10\%$}
    & Fast-adpgicp & 24.929 & 3.232 & 4.805 & 1.140 & 0.107 & 0.023 \\
    & GeoTransformer & 8.320 & 2.245 & 2.880 & 0.790 & 0.040 & 0.020 \\
    & Ours & \textbf{4.743} & \textbf{1.117} & \textbf{1.637} & \textbf{0.393} & \textbf{0.028} & \textbf{0.014} \\
    \midrule
    \multirow{3}{*}{$20\%$}
    & Fast-adpgicp & 17.775 & 3.787 & 4.010 & 1.261 & 0.047 & 0.024 \\
    & GeoTransformer & 6.564 & 1.999 & 2.256 & 0.709 & 0.034 & 0.018 \\
    & Ours & \textbf{4.570} & \textbf{1.832} & \textbf{1.573} & \textbf{0.442} & \textbf{0.027} & \textbf{0.010} \\
    \midrule
    \multirow{3}{*}{$30\%$}
    & Fast-adpgicp & 49.638 & 2.971 & 9.764 & 1.031 & 0.155 & 0.017 \\
    & GeoTransformer & 10.066 & 2.374 & 3.421 & 0.747 & 0.045 & 0.021 \\
    & Ours & \textbf{8.326} & \textbf{1.482} & \textbf{2.837} & \textbf{0.444} & \textbf{0.043} & \textbf{0.014} \\
    \midrule
    \multirow{3}{*}{$40\%$}
    & Fast-adpgicp & 65.932 & 2.641 & 14.636 & 0.889 & 0.160 & 0.023 \\
    & GeoTransformer & 10.814 & 2.392 & 3.665 & 0.765 & 0.048 & 0.020 \\
    & Ours & \textbf{6.437} & \textbf{1.950} & \textbf{2.194} & \textbf{0.425} & \textbf{0.033} & \textbf{0.018} \\
    \midrule
    \multirow{3}{*}{$50\%$}
    & Fast-adpgicp & 74.038 & 4.517 & 15.518 & 1.590 & 0.196 & 0.026 \\
    & GeoTransformer & 13.167 & 3.708 & 4.472 & 1.262 & 0.057 & 0.024 \\
    & Ours & \textbf{8.459} & \textbf{1.996} & \textbf{2.731} & \textbf{0.350} & \textbf{0.052} & \textbf{0.013} \\
    \bottomrule
  \end{tabular}
  }
\end{table}

The dropout experiment further demonstrates that the intensity cue provides additional reliability information for sparse point clouds, helping the registration remain stable when local geometric density is reduced.

\begin{figure}[t]
	\centering
	\includegraphics[width=1\linewidth]{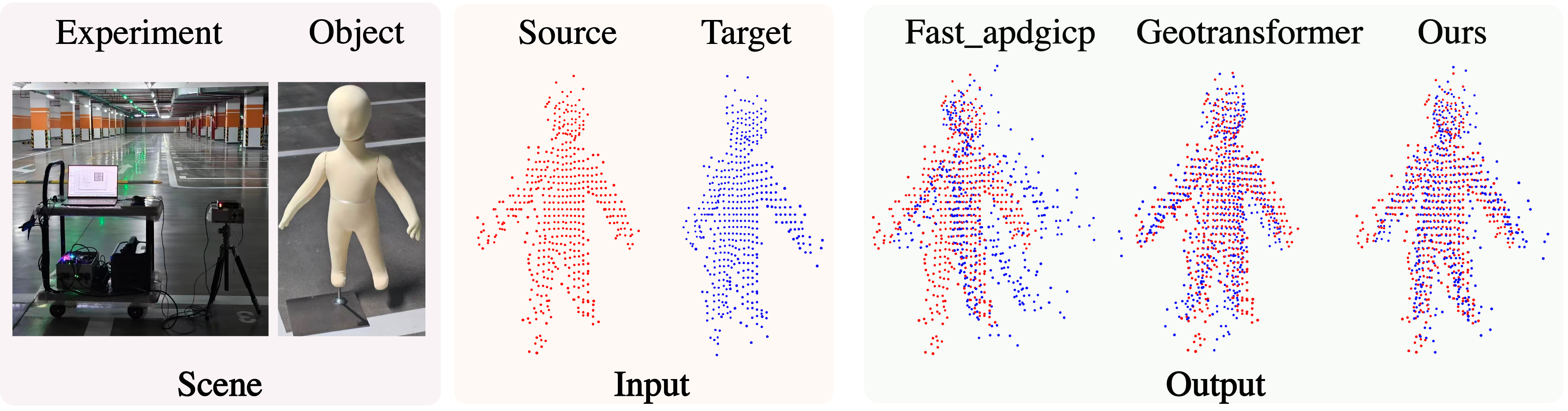}
	\caption{Qualitative registration comparison on real single-photon data acquired at a stand-off distance of about 80~m from two viewpoints. Fast-adpgicp incorrectly aligns body parts across different people, and GeoTransformer can produce a globally flipped registration without intensity information. The proposed method gives a more physically plausible orientation and more consistent local alignment. }
	\label{fig:realdata}
\end{figure}

\subsection{Results on the real dataset}\label{sec:real_results}

 The real-data experiment is used as a qualitative failure-mode comparison. As shown in Fig.~\ref{fig:realdata}, Fast-adpgicp is sensitive to local false overlaps caused by noisy body regions, while GeoTransformer can produce a flipped alignment when geometry-only correspondences are dominated by noisy but spatially plausible structures. By contrast, the proposed method gives a more physically plausible orientation and more consistent local alignment around the torso and arm regions. This observation is consistent with the synthetic failure cases and suggests that the intensity cue and OT-based unmatched-point rejection help suppress unreliable correspondences in real single-photon measurements. 

\subsection{Ablation Study}\label{sec:ablation}

\subsubsection{Effect of input intensity mode}

 The ablation results on different intensity input modes are first given in Table~\ref{tab:intensity_mode}. It is found that the normalized intensity achieves lower RTE, RRE, and RMSE than sigmoid intensity and raw intensity. Compared with raw intensity, the RRE is reduced from $9.106^\circ$ to $4.570^\circ$, while the RTE and RMSE are reduced from $3.097$~cm and $0.043$~m to $1.573$~cm and $0.027$~m, respectively. Compared with sigmoid intensity, the normalized intensity also reduces the RRE from $7.846^\circ$ to $4.570^\circ$, and the RMSE from $0.038$~m to $0.027$~m. It further indicates that normalization preserves the useful relative intensity variation, thus avoiding the feature-scale imbalance caused by raw intensity and the over-compression caused by sigmoid mapping.  The median errors follow the same trend, with normalized intensity reducing the median RRE to $1.832^\circ$ and the median RMSE to $0.010$~m, which indicates that the improvement is not only caused by a few favorable cases.

\begin{table}[t]
\centering
\caption{Ablation study on different intensity input modes for correspondence learning. Mean and median errors are reported. }
\label{tab:intensity_mode}
\small
{
\setlength{\tabcolsep}{3pt}
\begin{tabular}{lcccccc}
\toprule
\textbf{Feature mode} & \multicolumn{2}{c}{\textbf{RRE ($^\circ$) $\downarrow$}} & \multicolumn{2}{c}{\textbf{RTE (cm) $\downarrow$}} & \multicolumn{2}{c}{\textbf{RMSE (m) $\downarrow$}} \\
\cmidrule(lr){2-3}\cmidrule(lr){4-5}\cmidrule(lr){6-7}
& \textbf{mean} & \textbf{median} & \textbf{mean} & \textbf{median} & \textbf{mean} & \textbf{median} \\
\midrule
Normalization intensity & \textbf{4.570} & \textbf{1.832} & \textbf{1.573} & \textbf{0.442} & \textbf{0.027} & \textbf{0.010} \\
Sigmoid intensity & 7.846 & 3.241 & 2.715 & 1.149 & 0.038 & 0.026 \\
Raw intensity & 9.106 & 3.272 & 3.097 & 1.131 & 0.043 & 0.025 \\
\bottomrule
\end{tabular}
}
\end{table}
	
\subsubsection{Effect of feature aggregation strategy}

 The ablation results on different feature aggregation strategies are further summarized in Table~\ref{tab:feature_aggregation}. It shows that the aggregation strategy of combining KPConv with XY-Grid in our model achieves the best registration performance. Compared with KPConv only, its RRE is reduced from $6.640^\circ$ to $4.570^\circ$, and the RMSE is reduced from $0.032$~m to $0.027$~m, indicating the XY-Grid constraint helps regularize the coarse Top-$K$ selection and reduces incorrect candidate correspondences caused by noisy local structures. Furthermore, using XY-Grid only leads to a much larger RRE of $27.420^\circ$ and RMSE of $0.103$~m, as it does not capture the local geometric features learned from neighboring points. Therefore, KPConv is complementary to XY-Grid that the former preserves local geometric descriptors while the latter enhances the constraints on the uncertain depth and intensity-related feature responses. Its median RRE and RTE also remain high at $29.388^\circ$ and $10.146$~cm, confirming that the grid constraint alone cannot replace local geometric feature extraction.

\begin{table}[t]
\centering
\caption{Ablation study on different feature aggregation strategies. Mean and median errors are reported. }
\label{tab:feature_aggregation}
\small
{
\setlength{\tabcolsep}{3pt}
\begin{tabular}{lcccccc}
\toprule
\textbf{Feature aggregation strategy} & \multicolumn{2}{c}{\textbf{RRE ($^\circ$) $\downarrow$}} & \multicolumn{2}{c}{\textbf{RTE (cm) $\downarrow$}} & \multicolumn{2}{c}{\textbf{RMSE (m) $\downarrow$}} \\
\cmidrule(lr){2-3}\cmidrule(lr){4-5}\cmidrule(lr){6-7}
& \textbf{mean} & \textbf{median} & \textbf{mean} & \textbf{median} & \textbf{mean} & \textbf{median} \\
\midrule
KPConv \& XY-Grid & \textbf{4.570} & \textbf{1.832} & \textbf{1.573} & \textbf{0.442} & \textbf{0.027} & \textbf{0.010} \\
KPConv only & 6.640 & 2.754 & 2.301 & 0.968 & 0.032 & 0.020 \\
XY-Grid only & 27.420 & 29.388 & 9.454 & 10.146 & 0.103 & 0.113 \\
\bottomrule
\end{tabular}
}
\end{table}

	\section{Conclusions}\label{sec6}
	
	 We have presented the GIC-Reg framework to accomplish the  pose-free pairwise registration task for single-photon multi-view sensing. To address the issues of background contamination, spatial sparsity, and view-dependent depth degradation in single-photon point clouds, it is configured with modules of intensity-guided preprocessing, joint geometry-intensity grid feature aggregation, global matching, and local ambiguity disambiguation to estimate inter-view transformations and support subsequent multi-view reconstruction. Moreover, numerical experiments show that it is robust to both background noise and spatial dropout. Notably, it reduces the RRE from $49.805^\circ$ for Fast-adpgicp and $9.109^\circ$ for GeoTransformer to $8.517^\circ$, and reduces the RMSE to $0.039$~m, under the most degraded background-noise, and also achieves a lower RRE of $8.459^\circ$ and RMSE of $0.052$~m under $50\%$ dropout. Finally, outdoor experiments show that intensity information helps suppress false overlap, avoids globally flipped alignments, and produces more reliable local matching in human-body regions at a stand-off distance of about 80~m. These results confirm that our model, incorporating photon intensity as a confidence cue, helps stabilize pose-free pairwise registration in single-photon point clouds.

  There are many interesting problems to be addressed in further study. For example, the angular difference between the current viewpoints is relatively small, and the intensity variation caused by different observation angles is not yet considered. It would also be interesting to examine the registration performance of our model under much more extreme environments.

	\begin{backmatter}
		\bmsection{Funding}
		This research is supported by the National Natural Science Foundation of China (U22B201225), the Aeronautical Science Foundation of China (2022Z073038001) and the Fundamental Research Funds for the Central Universities (22120260123).

		\bmsection{Disclosures}
		The authors declare no conflict of interest.

			\bmsection{Data availability}
			Data underlying the results presented in this paper are not publicly available at this time but may be obtained from the authors upon reasonable request.

			\par\medskip
			\noindent See Supplement 1 for supporting content.
			
			\end{backmatter}

			{
		\bibliography{reference}

\begin{thebibliography}{10}
\newcommand{\enquote}[1]{``#1''}

\bibitem{zhang2022large}
X.~Zhang, K.~Kwon, J.~Henriksson, \emph{et~al.}, \enquote{A large-scale
  microelectromechanical-systems-based silicon photonics lidar,}
  {\protect\JournalTitle{Nature}} \textbf{603}, 253--258 (2022).

\bibitem{preussler2019photonically}
S.~Preussler, F.~Schwartau, J.~Schoebel, and T.~Schneider,
  \enquote{Photonically synchronized large aperture radar for autonomous
  driving,} {\protect\JournalTitle{Optics Express}} \textbf{27}, 1199--1207
  (2019).

\bibitem{chang2019new}
Y.-P. Chang, C.-N. Liu, Z.~Pei, \emph{et~al.}, \enquote{New scheme of
  lidar-embedded smart laser headlight for autonomous vehicles,}
  {\protect\JournalTitle{Optics Express}} \textbf{27}, A1481--A1489 (2019).

\bibitem{shi2019photonic}
J.-W. Shi, J.-I. Guo, M.~Kagami, \emph{et~al.}, \enquote{Photonic technologies
  for autonomous cars: feature introduction,} {\protect\JournalTitle{Optics
  Express}} \textbf{27}, 7627--7628 (2019).

\bibitem{chan2025flash}
J.~Chan, X.~Deng, T.~L.-t. Lee, \emph{et~al.}, \enquote{Flash and
  point-and-shoot hybrid lidar by dmd-based solid-state diffractive beam and
  image steering,} {\protect\JournalTitle{Optics Express}} \textbf{33},
  19650--19663 (2025).

\bibitem{maeda2025expanding}
E.~E. Maeda, B.~Brede, K.~Calders, \emph{et~al.}, \enquote{Expanding forest
  research with terrestrial lidar technology,} {\protect\JournalTitle{Nature
  Communications}} \textbf{16}, 8853 (2025).

\bibitem{cheng2017generalized}
Z.~Cheng, D.~Liu, Y.~Zhang, \emph{et~al.}, \enquote{Generalized
  high-spectral-resolution lidar technique with a multimode laser for aerosol
  remote sensing,} {\protect\JournalTitle{Optics Express}} \textbf{25},
  979--993 (2017).

\bibitem{mao2022polarization}
S.~Mao, A.~Wang, Y.~Yi, \emph{et~al.}, \enquote{Polarization raman lidar for
  atmospheric correction during remote sensing satellite calibration:
  instrument and test measurements,} {\protect\JournalTitle{Optics Express}}
  \textbf{30}, 11986--12007 (2022).

\bibitem{yuan2025remote}
D.~Yuan, Y.~He, and D.~Pan, \enquote{Remote sensing of particulate organic
  carbon in the south china sea using airborne high-spectral-resolution lidar,}
  {\protect\JournalTitle{Optics Express}} \textbf{33}, 32560--32576 (2025).

\bibitem{scholes2024robust}
S.~Scholes, E.~Wade, A.~McCarthy, \emph{et~al.}, \enquote{Robust framework for
  modelling long range dtof spad lidar performance,}
  {\protect\JournalTitle{Optics Express}} \textbf{32}, 47735--47756 (2024).

\bibitem{fan2021unsupervised}
S.~Fan, S.~Liu, X.~Zhang, \emph{et~al.}, \enquote{Unsupervised deep learning
  for 3d reconstruction with dual-frequency fringe projection profilometry,}
  {\protect\JournalTitle{Optics Express}} \textbf{29}, 32547--32567 (2021).

\bibitem{xu2025photon}
F.~Xu, \enquote{Photon-number-resolving detection enables single-photon lidar
  approaching the standard quantum limit,} {\protect\JournalTitle{Light:
  Science \& Applications}} \textbf{14}, 206 (2025).

\bibitem{tachella2019real}
J.~Tachella, Y.~Altmann, N.~Mellado, \emph{et~al.}, \enquote{Real-time 3d
  reconstruction from single-photon lidar data using plug-and-play point cloud
  denoisers,} {\protect\JournalTitle{Nature Communications}} \textbf{10}, 4984
  (2019).

\bibitem{li2025noise}
H.~Li, K.~Zheng, R.~Ge, \emph{et~al.}, \enquote{Noise-tolerant lidar
  approaching the standard quantum-limited precision,}
  {\protect\JournalTitle{Light: Science \& Applications}} \textbf{14}, 138
  (2025).

\bibitem{liu2023compact}
H.~Liu, C.~Qin, G.~Papangelakis, \emph{et~al.}, \enquote{Compact all-fiber
  quantum-inspired lidar with over 100 db noise rejection and single photon
  sensitivity,} {\protect\JournalTitle{Nature Communications}} \textbf{14},
  5344 (2023).

\bibitem{lee2023caspi}
J.~Lee, A.~Ingle, J.~V. Chacko, \emph{et~al.}, \enquote{Caspi: collaborative
  photon processing for active single-photon imaging,}
  {\protect\JournalTitle{Nature Communications}} \textbf{14}, 3158 (2023).

\bibitem{tobin2021robust}
R.~Tobin, A.~Halimi, A.~McCarthy, \emph{et~al.}, \enquote{Robust real-time 3d
  imaging of moving scenes through atmospheric obscurant using single-photon
  lidar,} {\protect\JournalTitle{Scientific Reports}} \textbf{11}, 11236
  (2021).

\bibitem{Li2021}
Z.-P. Li, J.-T. Ye, X.~Huang, \emph{et~al.}, \enquote{Single-photon imaging
  over 200 km,} {\protect\JournalTitle{Optica}} \textbf{8}, 344--349 (2021).

\bibitem{cheng2024toward}
S.~Cheng, H.~Qi, Y.~Sun, \emph{et~al.}, \enquote{Toward quantum unmanned
  systems,} {\protect\JournalTitle{Science China. Technological Sciences}}
  \textbf{67}, 2277--2280 (2024).

\bibitem{li2024high}
X.~Li, J.~Liu, G.~Zhao, \emph{et~al.}, \enquote{High precision single-photon
  object detection via deep neural networks,} {\protect\JournalTitle{Optics
  Express}} \textbf{32}, 37224--37237 (2024).

\bibitem{Shin2015}
D.~Shin, A.~Kirmani, V.~K. Goyal, and J.~H. Shapiro, \enquote{Photon-efficient
  computational 3-d and reflectivity imaging with single-photon detectors,}
  {\protect\JournalTitle{IEEE Transactions on Computational Imaging}}
  \textbf{1}, 112--125 (2015).

\bibitem{Rayman2006}
R.~Rayman, K.~Croome, N.~Galbraith, \emph{et~al.}, \enquote{Long-distance
  robotic telesurgery: a feasibility study for care in remote environments,}
  {\protect\JournalTitle{The International Journal of Medical Robotics and
  Computer Assisted Surgery}} \textbf{2}, 216--224 (2006).

\bibitem{Zang2019}
S.~Zang, M.~Ding, D.~Smith, \emph{et~al.}, \enquote{The impact of adverse
  weather conditions on autonomous vehicles: How rain, snow, fog, and hail
  affect the performance of a self-driving car,} {\protect\JournalTitle{IEEE
  Vehicular Technology Magazine}} \textbf{14}, 103--111 (2019).

\bibitem{Li2020}
Z.-P. Li, X.~Huang, P.-Y. Jiang, \emph{et~al.}, \enquote{Super-resolution
  single-photon imaging at 8.2 kilometers,} {\protect\JournalTitle{Optics
  Express}} \textbf{28}, 4076--4087 (2020).

\bibitem{Li2020a}
Z.-P. Li, X.~Huang, Y.~Cao, \emph{et~al.}, \enquote{Single-photon computational
  3d imaging at 45 km,} {\protect\JournalTitle{Photonics Research}} \textbf{8},
  1532--1540 (2020).

\bibitem{dai2023long}
C.~Dai, W.-L. Ye, C.~Yu, \emph{et~al.}, \enquote{Long-range photon-efficient 3d
  imaging without range ambiguity,} {\protect\JournalTitle{Optics Letters}}
  \textbf{48}, 1542--1545 (2023).

\bibitem{hadfield2023single}
R.~H. Hadfield, J.~Leach, F.~Fleming, \emph{et~al.}, \enquote{Single-photon
  detection for long-range imaging and sensing,}
  {\protect\JournalTitle{Optica}} \textbf{10}, 1124--1141 (2023).

\bibitem{Shi2022}
H.~Shi, G.~Shen, H.~Qi, \emph{et~al.}, \enquote{Noise-tolerant bessel-beam
  single-photon imaging in fog,} {\protect\JournalTitle{Optics Express}}
  \textbf{30}, 12061--12068 (2022).

\bibitem{Zhang2022}
Y.~Zhang, S.~Li, J.~Sun, \emph{et~al.}, \enquote{Three-dimensional
  single-photon imaging through realistic fog in an outdoor environment during
  the day,} {\protect\JournalTitle{Optics Express}} \textbf{30}, 34497--34509
  (2022).

\bibitem{buller2023single}
G.~S. Buller, A.~McCarthy, A.~Maccarone, \emph{et~al.}, \enquote{Single-photon
  lidar in challenging imaging scenarios,} in \emph{Advanced Photon Counting
  Techniques XVII,}  (SPIE, 2023), p. PC125120A.

\bibitem{jiang2023long}
P.-Y. Jiang, Z.-P. Li, W.-L. Ye, \emph{et~al.}, \enquote{Long range 3d imaging
  through atmospheric obscurants using array-based single-photon lidar,}
  {\protect\JournalTitle{Optics Express}} \textbf{31}, 16054--16066 (2023).

\bibitem{maccarone2015underwater}
A.~Maccarone, A.~McCarthy, X.~Ren, \emph{et~al.}, \enquote{Underwater depth
  imaging using time-correlated single-photon counting,}
  {\protect\JournalTitle{Optics Express}} \textbf{23}, 33911--33926 (2015).

\bibitem{Maccarone2023}
A.~Maccarone, K.~Drummond, A.~McCarthy, \emph{et~al.}, \enquote{Submerged
  single-photon lidar imaging sensor used for real-time 3d scene reconstruction
  in scattering underwater environments,} {\protect\JournalTitle{Optics
  Express}} \textbf{31}, 16690--16708 (2023).

\bibitem{Katzschmann2018}
R.~K. Katzschmann, J.~DelPreto, R.~MacCurdy, and D.~Rus, \enquote{Exploration
  of underwater life with an acoustically controlled soft robotic fish,}
  {\protect\JournalTitle{Science Robotics}} \textbf{3}, eaar3449 (2018).

\bibitem{shangguan2023compact}
M.~Shangguan, Z.~Yang, Z.~Lin, \emph{et~al.}, \enquote{Compact long-range
  single-photon underwater lidar with high spatial--temporal resolution,}
  {\protect\JournalTitle{IEEE Geoscience and Remote Sensing Letters}}
  \textbf{20}, 1--5 (2023).

\bibitem{Rapp2017}
J.~Rapp and V.~K. Goyal, \enquote{A few photons among many: Unmixing signal and
  noise for photon-efficient active imaging,} {\protect\JournalTitle{IEEE
  Transactions on Computational Imaging}} \textbf{3}, 445--459 (2017).

\bibitem{liu2025point}
J.~Liu, G.~Zhao, L.~Liu, \emph{et~al.}, \enquote{Point upsampling networks for
  single-photon sensing,} {\protect\JournalTitle{arXiv:2508.12986}}  (2025).

\bibitem{wennoise}
Z.~Wen, W.~Dong, Z.~Zhang, \emph{et~al.}, \enquote{Noise-aware adaptation of
  pre-trained foundation models for single-photon image classification,}
  {\protect\JournalTitle{Trans. Mach. Learn. Res.}} \textbf{2026} (2026).
  {OpenReview forum qSnrIy6Ohb}.

\bibitem{yang2024pe}
X.~Yang, S.~Xiao, H.~Zhang, \emph{et~al.}, \enquote{Pe-rasp: range image
  stitching of photon-efficient imaging through reconstruction, alignment,
  stitching integration network based on intensity image priors,}
  {\protect\JournalTitle{Optics Express}} \textbf{32}, 2817--2838 (2024).

\bibitem{malik2023transient}
A.~Malik, P.~Mirdehghan, S.~Nousias, \emph{et~al.}, \enquote{Transient neural
  radiance fields for lidar view synthesis and 3d reconstruction,}
  {\protect\JournalTitle{Advances in Neural Information Processing Systems}}
  \textbf{36}, 71569--71581 (2023).

\bibitem{luo2025transientangelo}
W.~Luo, A.~Malik, and D.~B. Lindell, \enquote{Transientangelo: Few-viewpoint
  surface reconstruction using single-photon lidar,} in \emph{2025 IEEE/CVF
  Winter Conference on Applications of Computer Vision (WACV),}  (IEEE, 2025),
  pp. 8723--8733.

\bibitem{thomas2019kpconv}
H.~Thomas, C.~R. Qi, J.-E. Deschaud, \emph{et~al.}, \enquote{Kpconv: Flexible
  and deformable convolution for point clouds,} in \emph{Proceedings of the
  IEEE/CVF international conference on computer vision,}  (2019), pp.
  6411--6420.

\bibitem{zhang20234dradarslam}
J.~Zhang, H.~Zhuge, Z.~Wu, \emph{et~al.}, \enquote{4dradarslam: A 4d imaging
  radar slam system for large-scale environments based on pose graph
  optimization,} in \emph{2023 IEEE International Conference on Robotics and
  Automation (ICRA),}  (IEEE, 2023), pp. 8333--8340.

\bibitem{qin2023geotransformer}
Z.~Qin, H.~Yu, C.~Wang, \emph{et~al.}, \enquote{Geotransformer: Fast and robust
  point cloud registration with geometric transformer,}
  {\protect\JournalTitle{IEEE Transactions on Pattern Analysis and Machine
  Intelligence}} \textbf{45}, 9806--9821 (2023).

\bibitem{shin2017photon}
D.~Shin, J.~H. Shapiro, and V.~K. Goyal, \enquote{Photon-efficient
  super-resolution laser radar,} in \emph{Wavelets and Sparsity XVII,}  vol.
  10394 (SPIE, 2017), pp. 9--16.

\bibitem{wu2015modelnet}
Z.~Wu, S.~Song, A.~Khosla, \emph{et~al.}, \enquote{3d shapenets: A deep
  representation for volumetric shapes,} in \emph{Proceedings of the IEEE
  conference on computer vision and pattern recognition,}  (2015), pp.
  1912--1920.

\bibitem{chen2022deep}
Y.~Chen, G.~Yao, Y.~Liu, \emph{et~al.}, \enquote{Deep domain adversarial
  adaptation for photon-efficient imaging,} {\protect\JournalTitle{Phys. Rev.
  Appl.}} \textbf{18}, 054048 (2022).

\end{thebibliography}
	}

\end{document}